\documentclass[12pt]{article}
\usepackage{geometry}
\usepackage{authblk}
\usepackage{hyperref}
\usepackage{appendix}
\usepackage{graphicx, wrapfig}
\graphicspath{{C:/Users/Rem/Documents/Documents/PhD/Thesis/Decentralized Token Economy Theory (DeTEcT)/Images/}}
\usepackage{amssymb}
\usepackage{amsmath}
\usepackage{amsthm}
\usepackage{esvect}
\usepackage{mathtools}
\usepackage{epigraph}
\usepackage{fancyhdr}
\usepackage{xcolor}
\usepackage{caption}
\newtheorem{definition}{Definition}
\newtheorem*{remark}{Remark}
\newtheorem*{axiom}{Axiom}

\title{\Large \textbf{Decentralized Token Economy Theory (DeTEcT)}\\ \large Token pricing, stability and governance for token economies}
\author{R. Sadykhov, Dr.G. Goodell, Dr.D. de Montigny, Dr.M. Schoernig and Prof.P. Treleaven}
\affil{University College London}
\date{June 2023}

\begin{document}

\begin{titlepage}
\maketitle
\thispagestyle{empty}
\begin{abstract}
This paper presents a pioneering approach for simulation of economic activity, policy implementation, and pricing of goods in token economies. The paper proposes a formal analysis framework for wealth distribution analysis and simulation of interactions between economic participants in an economy. Using this framework, we define a mechanism for identifying prices that achieve the desired wealth distribution according to some metric, and stability of economic dynamics.\par
The motivation to study \emph{tokenomics} theory is the increasing use of tokenization, specifically in financial infrastructures, where designing token economies is in the forefront. \emph{Tokenomics} theory establishes a quantitative framework for wealth distribution amongst economic participants and implements the algorithmic regulatory controls mechanism that reacts to changes in economic conditions.\par
In our framework, we introduce a concept of tokenomic taxonomy where agents in the economy are categorized into agent types and interactions between them. This novel approach is motivated by having a generalized model of the macroeconomy with controls being implemented through interactions and policies. The existence of such controls allows us to measure and readjust the wealth dynamics in the economy to suit the desired objectives.
\end{abstract}
\end{titlepage}

\section{Introduction}
\emph{Tokenomics} (token economics) is a theory that embodies distribution of tokenized goods and services between economic participants. The purpose of tokenomics is to analyse an existing token economy, or to help design new token economies. Tokenomics requires a policy setting mechanism for pricing, risk management, stability analysis, etc.\par
\emph{Tokenization} is a process of issuing tokens that represent a digital asset, where the rules and behaviors governing the asset are controlled by smart contracts with the program outcomes stored on a blockchain \cite{LawOfTokenomics}. Asset tokenization is typically performed on existing blockchains (e.g., ERC20 tokens on Ethereum blockchain). A token has an identifier, a set of associated defining properties (rules and behaviors) and is secured cryptographically. There exist different types of tokens according to International Organization for Standardization \cite{ISOVocabulary}, specifically:
\begin{itemize}
	\item \textbf{Fungible Token} - token that is capable of mutual substitution among individual units.
	\item \textbf{Non-Fungible Token (NFT)} - token that is not capable of mutual substitution among individual units.
	\item \textbf{Security Token} - token with speciific characteristics that meets the definition of financial instrument or other investment instrument under applicable legislation in the relevant jurisdiction.
	\item \textbf{Utility Token} - token that can be used by its owner to receive access to goods or services.
\end{itemize}
We consider the standard definition for economics, which is the study of efficient allocation of scarce resources \cite{ISOVocabularyEconomics}, or as a study of wealth \cite{WealthOfNations}. Acting under the assumption that tokens (or any unit of currency) are used as the storage of wealth, and the medium for valuations and transactions, we conclude that tokenomics is the study of efficient allocation of wealth represented by tokens.\par
The major questions that tokenomics aims to address is how does an economic system (e.g., blockchain protocol, smart contract) provisions and allocates scarce resources (i.e., coins and tokens), how does this system interact with the external economic systems (e.g., market capitalization of a cryptocurrency), how do agents in an economy behave (e.g., demand, saving propensity), and what is the efficiency of these processes.\par
To achieve controlled and regulated distribution of tokens, a tokenomics theory is needed. This theory must have a framework for simulation of token distribution dynamics which facilitates the construction of a control mechanism to achieve the desired dynamics. Examples of implementation where this control mechanism can be utilized are the programs and applications that run in Web 3.0 \cite{Web3.0TokenizationAndDeFi} :
\begin{itemize}
	\item \textbf{Decentralized Applications (DApps)} - application that runs on a decentralized system \cite{ISOVocabulary}.
	\item \textbf{Decentralized Autonomous Organization (DAO)} - an entity structure in which tokenholders participate in the management and decision-making of an entity \cite{DAO}.
	\item \textbf{Decentralized Finance (DeFi)} - financial services that use remove third-parties and centralized institutions from transactions \cite{DeFi}.
	\item \textbf{Game Finance (GameFi)} - blockchain enabled games that offer economic incentives to play them \cite{GameFi}.
\end{itemize}
These applications require control over the policy setting mechanism that distributes tokenized assets and mitigates risks associated with extreme conditions of economic dynamics (e.g., recession, inflation, price volatility, etc.), as well as performs pricing of goods and services offered in these token economies.\par
We propose Decentralized Token Economy Theory (DeTEcT) as a theory that is able to model the distribution of tokenized assets, perform stability analysis of token economies and price tokenized goods and services. Simultaneously, DeTEcT is capable of highlighting systematic weaknesses in token economies as well as simulating behavior of participants in different scenarios.\par
DeTEcT's simulation framework is an agent-based dynamical system that models wealth distribution between macroeconomic agents in a token economy. The dynamical system has two main applications:
\begin{itemize}
	\item \textbf{Forward Propagation} - Prices, $p$, and demands (transaction requests), $d$, over a specified period $\Delta t$ are used to simulate the wealth distribution dynamics. This approach is used to demonstrate hidden patterns in wealth distribution from historical data and can be applied to simulate a possible future wealth distribution.
	\item \textbf{Inverse Propagation} - Dynamical system is solved based on the desired final state of wealth distribution to obtain parameters that must be maintained to achieve the desired wealth distribution. These parameters configure rate equations that define balance between prices, $p$, and demands (transaction requests), $d$. Rate equations define all possible solutions for prices in terms of the transaction requests.
\end{itemize}
The aim of this paper is to introduce a formal analysis framework for the dynamics of wealth distribution in token economies where policies (e.g., transaction fees, agent restrictions, money supply) are treated as variables that control the wealth distribution dynamics in a predictable way.\par
To achieve this, we introduce definitions that enable us to construct a compartmental model that exhibits predictable wealth distribution dynamics. Moreover, we consider that a token economy has a mechanism that controls money supply, transaction fees, and sets other economic policies (e.g., transaction restrictions). No agent in the economy can become this mechanism, but there can be some governance over this mechanism that regulates its actions. In general, we refer to this mechanism as Control Mechanism and consider it to be an agent in our framework.\par
Section \ref{Applications} discusses use cases for the framework and what range of problems can be solved using it, as oppose to showcasing the exact results. In that section, we introduce a separate scope for the simulations, where we demonstrate how the theoretical framework introduced in section \ref{MathematicalFramework} can be applied to modelling the wealth distribution in a closed economy. For the purpose of the demonstration we will set up a fictional economy with a simple interactions taxonomy to be used in the examples.\par
When defining the mechanism for identification of prices, we limit the discussion of pricing goods and services in a token economy to the case where all prices are set by the Control Mechanism and the economy resembles the command economy. This assumption is not the constraint of the proposed framework, but rather a simplification for the purpose of the paper.

\section{Related Work}
\subsection{Wealth Distribution Models}
To model macroeconomy and its governance, wealth distribution models are often used to describe money supply, taxation and other regulatory controls, which is a common research topic in classical economics as discussed in sections \ref{ApproachesToModelWealthDistribution} and \ref{ApproachesToControlWealthDistribution}. Multiple approaches have been proposed to model a wealth distribution between agents in a closed economy and control its dynamics. We can broadly categorize these approaches into modelling and controlling the wealth distribution. The objective of the former is to simulate a real-world wealth distribution between agents, whilst the latter is used to implement a control mechanism to get a desired wealth distribution.
\subsubsection{Approaches to Model Wealth Distribution}
\label{ApproachesToModelWealthDistribution}
In academic literature, wealth distribution models often assume that wealth reallocation happens in the form of random events so the statistical agent-based models are used for simulations. Examples of such models are presented by Dragulescu et al. \cite{StatisticalMechanicsOfMoney}, Chakraborti et al. \cite{StatisticalMechanicsOfMoneySavingPropensity} and Chatterjee et al. \cite{MoneyInGasLikeMarkets}, where wealth distribution is simulated in a similar way to interacting gas molecules in a closed container. These approaches are using random interactions in an economy (much like random collisions of gas molecules) to simulate the equilibrium in wealth distribution.\par
Statistical properties of these models are convenient to describe distribution of wealth using Pareto law \cite{KineticTheoryModelsForTheDistributionOfWealth} and measure the relaxation time for the wealth distribution to return to its equilibrium \cite{RelaxationInStatisticalManyAgentEconomyModels}. However, these models do not help with the implementation of controls over the wealth distribution dynamics in a closed economy setting since they primarily focus on the equilibria states of microeconomic systems.
\subsubsection{Approaches to Control Wealth Distribution}
\label{ApproachesToControlWealthDistribution}
For the implementation of controls in economic systems, control theory is often used where a system of equations describing the economic system is considered along with a criterion function that stops the algorithm when an optimal state is reached. Control theory can be used in deterministic and statistical settings. The main methods for implementing control theory for economic systems are summarized in \emph{``Applications of Control Theory to Macroeconomics''} \cite{ApplicationsOfControlTheoryToMacroeconomics} with implementation examples.\par
Note that these implementations use the agent-based modelling technique and consider a vector of state variables, $x_{t}$, and a vector of control variables, $u_{t}$. The control variables are the settings controlling the economic mechanism.\par
This approach is applicable to modelling a wealth distribution where the states are the agents' wealths that dynamically evolve in the economy due to interactions between agents. The control variables are the policies introduced and their impact on the state dynamics and the final outcome can be measured.\par
\subsubsection{Wealth Distribution in Token Economies}
Most of the existing academic work on wealth distribution in token economies is concerned with the centralization of wealth. Centralization, or concentration, of wealth in an economy can lead to different economic threats, such as iFish attack in 2018 \cite{IFish}, where the whale transactions (i.e., transactions with high fee) on the Ethereum blockchain caused an increase in the transaction fees across the network by 37\%, which led to the censorship of transactions with smaller fees and reduced the throughput in the economy (less transactions could go through). Another threat specific to the token economies is the centralization of governance in the economy, as token economies employ distributed ledgers for record keeping of transaction history and the proof of ownership over an asset. In most token economies, the tokens themselves act not only as a transaction medium, but also as a voting mechanism, which establishes the consensus over the policy changes to be implemented to the ledgers. If an economic participant was to gain ownership over a significant proportion of tokens, the governance of the system will be prone to malicious actions of this participant. Due to the different risks associated with the centralization of wealth, it is natural to enquire how centralized different token economies are in terms of their wealth distribution.\par
The wealth distribution of cryptocurrencies has been examined in \emph{Characterizing Wealth Inequality in Cryptocurrencies} \cite{CharacterizingWealthInequality} where Bitcoin, Ethereum and their respective forks have been studied. The process employed by Sai et al. \cite{CharacterizingWealthInequality} uses the Gini coefficient \cite{Gini} and Nakamoto index (minimum number of nodes required to disrupt the blockchain network; commonly, number of addresses whose combined wealth exceeds 51\% of total supply is used) to measure the inequality of wealth distribution. The study itself, involved the construction of a single data structure using the ETL (Extract, Transform, and Load) technique \cite{ETLProcess} to store transaction data across all studied token economies in one database. The study also looked at how do policy changes in Bitcoin Improvement Proposals (BIP) \cite{BIP} affect the wealth distribution in Bitcoin, and specifically the centralization of wealth.\par
The research of Sai et al. \cite{CharacterizingWealthInequality} summarized that the major contributor to wealth centralization is the market capitalization of the cryptocurrency, where Ethereum and Bitcoin seem to be more decentralized than their forks (e.g., Litecoin, Ethereum Cash). Another finding was that some BIPs seemed to have a significant impact on wealth distribution, with the value of Gini index for Bitcoin falling on multiple occasions after a BIP was implemented (e.g., when BIP42 introduced the maximum supply of Bitcoins). The research also reports that most cryptocurrencies start with a relatively centralized distribution of tokens, but over time the wealth distribution gets more dispersed causing the drop in the decentralization metrics where it stabilizes around the same levels as the wealth distribution in standard economies (e.g., the median Gini coefficient across countries' economies is 0.73, while the median across cryptocurrencies is 0.71 as reported in Janruary 2021).\par
Similar study has been performed by Ku\'smierz at al. \cite{HowCentralizedIsDecentralized}, where only the wealthiest addresses in a token economy were considered (i.e., case studies with top 30, 50, and 100 wealthiest addresses) to study the decentralization of governors or addresses that can affect the governance of the protocol. Along with Gini coefficient and Nakamoto index, Shannon entropy and Zipf coefficient (i.e., the parameter of the Zipf distribution) were considered as the decentralization metrics. The study demonstrated the trend of cryptocurrencies to have high degree of centralization closer to the genesis block or cryptocurrency launch, with the wealth being distributed in a more decentralized manner (at least between the governors) as the cryptocurrency matures. These results confirm the findings of Sai et al. \cite{CharacterizingWealthInequality}, and suggest that there might exist a natural equilibrium of wealth distribution in token economies.\par
\subsection{Tokenomics Models}
A number of tokenomic models have been proposed to deal with the specific problems arising in tokenomics (e.g., fee optimization, economic stability, etc.). In this section we will list a few tokenomics approaches and theories that we believe are of the most relevance to this paper. However, we must also mention that there exists a number of proprietary solutions that aim to optimize specific parts of a blockchain protocol such as the Gauntlet Protocol \cite{Gauntlet} that optimizes parameters of a protocol to achieve capital efficiency, fee maximization, etc.\par
In \emph{A Control Theoretic Approach to Infrastructure-Centric Blockchain Tokenomics}, Akcin et al. \cite{AControlTheoreticApproachToTokenomics} propose a model of control system implemented for blockchain through a mechanism called "Central" Treasury. This mechanism controls the supply of tokens in circulation by readjusting token payments to participants in a token economy and token buy-backs in order to achieve a stable token price. This study also introduces the use of dynamical systems for modelling the activity of the Central Treasury, and introduces a control theoretic approach to determine the parameters for the system to behave in a desired way.\par
To investigate the blockchain governance structure and the preference of the economic participants, Lee et al. \cite{AnAnalysisOfBlockchainGovernance} have devised a theory where new policies are suggested by proposers and the adoption of that policy is examined from the theoretical perspective. This theory specifically highlights how the governance of the blockchain influences the policy decisions and how can the governance structures be compared to examine what effects they have within a token economy (i.e., Bitcoin vs Ethereum governance).\par
Aside from governance, token economies demonstrate structural differences such as the existence of maximum supply or the fungibility of tokens. These structural differences of different token economies are examined in \emph{Tokenomics and Blockchain Tokens: A Design-Oriented Morphological Framework} by Freni et al. \cite{TokenomicsAndBlockchainTokens}, where the authors compare existing classification frameworks for token economies and based on that review derive a more complete classification framework. The proposed morphological token classification framework is useful for identifying different features implemented in token economies.\par
\subsection{Simulation Engines}
In section \ref{Applications} we will demonstrate the applications of our framework and scenarios in which it can be applied. This requires an engine for simulating the evolution of the wealth distribution dynamics.\par
Here we review two engines that were designed to run simulations of large systems. We look at the MMI simulator \cite{MacroMicroInterlockedSimulator} and its application as it gives us an insight on how the concurrent integration of macro- and micro-subsystems has been done before, while we review the World3 simulator \cite{TheLimitsToGrowth} as it demonstrates the implementation of the dynamical system modelling for large systems.
\subsubsection{Macro-Micro Interlocked (MMI) Simulator}
The MMI simulator proposed by Sato \cite{MacroMicroInterlockedSimulator} is a simulation engine that uses interrelated subsystems to run concurrent simulations of these subsystems. In \emph{``Macro-Micro Economic System Simulation''} \cite{MacroMicroEconomicSystemSimulation}, the engine was implemented via macro and micro subsystems to model an economy of a country, including its internal market and the trading interactions with other countries. Macro subsystem used Dornbusch model \cite{DornbuschModel}, which comprises the goods market, the money market and the asset market. Micro subsystem is implemented via the Firms-Labour model \cite{MicrosimulationModellingOfTheCorporateFirms}, where agents in the economy are used to simulate labour inside the economy.\par
The two subsystems have interlocked input-output format: the state of the micro model is fed into the macro model, and the state of the macro model is fed into the micro model. Taxation and money supply parameters in Dornbusch model are used to control the macro subsystem and their impact can be studied on the micro subsystem.\par
The approach this simulation engine takes is very flexible as different models can be used as subsystems, but the constraint on these models is that their input-output must be interlocked. One further observation is that the objective of this simulator is not to find the optimal control variables, but rather to simulate the dynamics of the chosen subsystem models.
\subsubsection{World3 Simulator}
World3 \cite{TheLimitsToGrowth} is a simulation engine for evolution of the Earth's environment and studying the dynamic interactions between features such as population, industrial growth, food production and pollution. The simulation engine, implemented separately by Vanwynsberghe \cite{PyWorld3} and Legavre \cite{MyWorld3}, uses a dynamical system of equations to model the interactions between the features.\par
The motivation for the introduction of the dynamical system in this simulation engine is to have easy integration of real-world empirical data through the parameters of the dynamical system. The drawback of using this method is the inaccuracies in predictions due to unaccounted shocks (e.g., technological breakthrough) that can change the dynamics of the system altogether.

\section{Methodology}
\subsection{General Approach}
Our objective is to build a framework to simulate the wealth distribution among agents in a token economy, and to enable the implementation and testing of algorithmic policy-making within this framework. For these purposes, we adopt the methodology from \emph{``Applications of Control Theory to Macroeconomics''} by Kendrick \cite{ApplicationsOfControlTheoryToMacroeconomics} to build the framework for simulations, where there exists a criterion function (e.g., economic metric, weighted economic metrics) and the system of equations (e.g., equations of wealth distribution dynamics). In this context, the parameters of our model are used to optimize the wealth distribution in a token economy according to a choice of metrics.\par
For the system of equations to be used in the control theory setting, we chose a similar method to Ross \cite{SIRModel}, where a bilinear dynamical system is employed to model the interactions between agents. It is a virology model used to simulate infection dynamics in population, but instead of population we will be using wealth as the quantity that gets redistributed between agents that are compartmentalized into agent types. In this paper we will be taking the deterministic approach t modelling, but we will also make remarks on adapting the framework to the probabilistic setting.\par
\subsection{Definitions Of Terms}
DeTEcT is a tokenomics theory that studies macro and microeconomic effects for policy setting based on behaviors of different participants as well as responses to changes in the economic climate (e.g., recession, inflation). DeTEcT uses dynamical system to model the interactions between different economic participants via interaction rates that are used to describe the dynamics of token distribution. By introducing a theoretical approach to studying tokenomics, we aim to minimize financial risks and optimize the performance of token economies for their set purposes.\par
The advantage of tokenized assets is the availability and completeness of historic transaction data on the ledger. This allows for ``complete'' modelling as there are no hidden cashflows.\par
To proceed, we must impose an assumption on time.\par
\begin{axiom}[Discrete Time]
	Time, $t\in T$, is discrete, with a constant interval $\Delta t$, such that
\begin{equation}
	T = \cup^{m}_{j=0}\{t_{j}\}, \quad m\in\mathbb{N}_{0},
\end{equation}
where $t_{0}$ marks the start of the simulation, and $t_{m}$ marks the end.
\end{axiom}
For technical purposes, we need an initialization step before $t_{0}$, which we call $t_{initial}$, during which an initial wealth allocation is made and the simulation is set up. Having imposed the assumption on time in our framework, we introduce the definitions fundamental to DeTEcT.\par
\begin{remark}
	Before describing the theoretical framework, we would like to elaborate on the sequence of operations performed at each timestep. From timestep $t$ to timestep $t+\Delta t$ we collect transaction data; at $t+\Delta t$ we first apply changes to policies we announced in previous steps, then we process all the actions performed by agents in the chronological order throughout the latest time interval $\Delta t$.
\end{remark}
We begin by introducing individual agents that will generate the economic activity in a token economy, much like individual participants do in real-world economies.\par
\begin{definition}
	\textbf{Set of all agents}, $\Lambda$, is a finite set of all users that participate in the economic activity. The total number of agents is the cardinality of $\Lambda$ and may vary in time as some agents leave or join.
\end{definition}
Each agent has a wealth that can be used to transact with other agents for goods and services.\par
\begin{definition}
\label{IndividualWealthFunction}
	\textbf{Individual wealth function} is a mapping $f:\Lambda\times T\rightarrow \mathbb{R}_{\geq0}$ to the aggregate value of wealth an individual agent holds at time $t\in T$.
\end{definition}
Note that the range of $f$ is $\mathbb{R}_{\geq0}$. We motivate the choice of this range by stating that most token economies do not support creation of credit which is why we assume that the token values are always non-negative.\par
Now that there exist individual agents, they can be categorized into agent types (e.g., households, producers, investors, etc.) based on some predefined features.\par
\begin{definition}
	\textbf{Agent type}, $\mathfrak{A}$, is a set of features called qualifying criteria, that categorize agents in $\Lambda$ into disjoint sets.
\end{definition}
Note that for a given timestep $t\in T$, an individual agent has a unique agent type, but for different timesteps, $t\neq t^{\prime}$, an agent can change its agent type (e.g., a consumer becoming a producer).\par
Given such categorization we unite agents into agent categories which are the sets of agents of the same agent type, where the union of agent categories is disjoint. The reason we introduce this limitation is to avoid double counting of agent's wealth in our model, and because from practical perspective we can always shrink the time interval $\Delta t$ until every agent interacts as a specific agent type only once per $\Delta t$ (we assume that there is always going to be at least some microscopic latency between interactions that an agent performs).\par
\begin{definition}
\label{AgentTypeSet}
	\textbf{Agent category}, $A$, is a finite set of individual agents of the same agent type $\mathfrak{A}$. The cardinality of a given agent category, $|A|$, is not constant in time as individual agents are free to choose to be affiliated with the agent type $\mathfrak{A}$ or not to be.
\end{definition}
We can now define a set of all agent categories that exist in a token economy through a pseudo-partition to allow for a more convenient set notation.\par
\begin{definition}
\label{PseudoPartitionOfAgentTypes}
	\textbf{Pseudo-partition of agent categories}, $E_{t}$, is a partition of all agents in $\Lambda$  into agent categories $A\in E_{t}$ at time $t\in T$,
	\begin{equation}
		\forall a\in\Lambda, \exists! A\in E_{t} \text{ such that } a\in A,
	\end{equation}
where $A=\emptyset$ if no agent is of agent type $\mathfrak{A}$. The cardinality of the pseudo-partition, $\vert E_{t}\vert$, is the total number of agent types and is constant with respect to $t$.
\end{definition}
Given the definitions \ref{AgentTypeSet} and \ref{PseudoPartitionOfAgentTypes}, we can characterize the set of all agents $\Lambda$ in terms of the agent categories $A_{j}\in E_{t}$ as
\begin{equation}
	\cup^{n}_{j=0}A_{j}=\sqcup^{n}_{j=0}A_{j}=\Lambda,
\end{equation}
where $n=\vert E_{t}\vert$ is the number of agent types in $E_{t}$ and $\sqcup$ is the disjoint union operator.\par

Each agent category must have an aggregate wealth we can use to model a token economy.\par
\begin{definition}
\label{WealthFunction}
	\textbf{Wealth function} is a mapping $F:\mathcal{P}(\Lambda)\times T\rightarrow\mathbb{R}_{\geq0}$ which is the aggregate value of wealth of a finite set $\mathcal{A}\in\mathcal{P}(\Lambda)$ at time $t$ such that
	\begin{equation}
		F(\mathcal{A},t) = \sum_{a\in\mathcal{A}}f(a,t),
	\end{equation}
where f is the individual wealth function as per definition \ref{IndividualWealthFunction}, and $\mathcal{P}(\Lambda)$ is the power set of $\Lambda$. For an agent category $A$, $F(A,t)$ is known as its wealth or holdings.
\end{definition}
To create a dynamical system, we require interactions and rotations between different agent categories.\par
\begin{definition}
	\textbf{Interaction}, $\xi$, between agents $a\in A$ and $a^{\prime}\in A^{\prime}$ is the exchange of goods and services that may result in wealth and resource redistribution, with $\xi$ a mapping
	\begin{equation}
		\xi:A\times A^{\prime}\times T\times I_{AA^{\prime}}\rightarrow\mathbb{R}, \quad A,A^{\prime}\in E_{t}, I_{AA^{\prime}}\subseteq I,
	\end{equation}
where $t\in T$ is the time, $s\in\mathbb{R}$ is the amount of wealth that was moved from agent $a$ to agent $a^{\prime}$ due to the interaction, and $I_{AA^{\prime}}$ is the set of all possible interactions between agents in agent categories $A$ and $A^{\prime}$. $I$ is the set of all possible interactions between agents across all agent categories,
	\begin{equation}
		I = \cup_{A,A^{\prime}\in E_{t}}I_{AA^{\prime}}.
	\end{equation}
 $s$ is referred to as the transaction quantity. The interactions with $s\neq0$ are called cashflow interactions or transactions as they result in wealth redistribution; the interactions with $s=0$ are called cashless interactions.
\end{definition}
Note that the negative $s$ of a cashflow interaction means that the wealth is redistributed in the direction from $a^{\prime}\in A^{\prime}$ to $a\in A$.\par
\begin{definition}
	\textbf{Rotation}, $r$, of agent $a\in\Lambda$ between $A,A^{\prime}\in E_{t}$ is the change of agent's $a$ affiliation from $A$ to $A^{\prime}$ such that
	\begin{equation}
		r(a, A^{\prime})\in A^{\prime} \quad \text{ for } a\in A.
	\end{equation}
\end{definition}
The motivation behind introducing rotations is the hard constraints we impose on the consumption of wealth. All interactions in our framework are defined through the interactions taxonomy (list of all possible interactions and agent categories that can participate in them), so for an agent to purchase a specific good (defined as an interaction), the agent must be affiliated to an agent category that can perform this interaction. Rotations are introduced for agents to facilitate the change of their agent categories and therefore, get access to different interactions (goods and services).\par
We assume that in a token economy there exists a maximum supply, $M$, which is the aggregate value of all tokens in the economy (including the promise of minting a specific amount of new tokens in the future). This allows us to define the distribution of wealth among the agent categories for every time $t$.\par
\begin{definition}
	\textbf{Wealth distribution}, $(F(A_{1},t),...,F(A_{n},t))^{T}\in\mathbb{R}^{n}$, is the distribution of wealth between different agent categories at time $t\in T$, such that
	\begin{equation}
		\sum_{A\in E_{t}}F(A,t)=M \quad \forall t\in T,
	\end{equation}
where $F$ is the wealth function, and $M$ is the time-independent maximum supply. The wealth distribution is achieved through interactions and rotations of individual agents between agent categories. The maximum supply $M$ does not have to be a constant in time, and instead can be time-dependent, $M(t)$.
\end{definition}
In this paper we use constant $M$ since the case of $M(t)$ adds complexity and has to be considered separately.\par
In a context of a token economy, we also define a special agent category called control mechanism that will perform the task of managing the economy.\par
\begin{definition}
	\textbf{Control mechanism}, $B\in E_{t}$, is a required agent category for a token economy and it consists of only one agent, $\vert B\vert=1$. The control mechanism is responsible for minting, distributing and burning tokens, as well as implementing other economic policies.
\end{definition}
Given the maximum supply $M$ and the control mechanism, $B\in E_{t}$, we define the circulation.\par
\begin{definition}
	\textbf{Circulation}, $S:T\rightarrow\mathbb{R}_{+}$ is a mapping to the wealth that is distributed between all agent categories except the control mechanism at time $t\in T$,
	\begin{equation}
	\label{CirculationEquation}
		S(t) = M-F(B,t).
	\end{equation}
\end{definition}
\subsection{Tokenomic Taxonomy}
Tokenomic taxonomy is the collection of unique interactions and rotations that agent categories can perform along with the counterparties for every possible interaction. It is a crucial component of tokenomics theory as it allows us to understand the potential cashflows and how the tokenomic dynamics evolves in time.\par
The tokenomic taxonomy does not necessarily identify every unique participant of the token economy, instead it categorizes participants into agent categories that have their own set of interactions that they can perform.\par
We state that an individual user (agent) may be eligible for different agent categories as that agent satisfies qualifying criteria for multiple agent types. DeTEcT allows agents in the economy to change roles by moving their tokens to a different agent category via rotations without registering a transaction.

\section{DeTEcT - Analytical Framework}
\label{MathematicalFramework}
\subsection{Dynamical System (Two Agent Categories)}
\label{DynamicalSystem}
Using the interactions and rotations listed in the tokenomic taxonomy we can simulate the dynamics of the token economy.\par
We propose a framework with a dynamical system at its core, as defined by Abraham et al. in \emph{``Foundations of Mechanics''} \cite{FoundationsOfMechanics}, which is a bilinear dynamical system similar to the model defined by Ross \cite{SIRModel}. Bilinear dynamical systems may not have unique solutions, and only some types of bilinear systems have the proof of existence of solution, which are outlined by Mohler et al. \cite{BLSAndApplications}, \v{C}elikovs\'{y} et al. \cite{BilinearSystemsAndChaos} and Johnson et al. \cite{SolutionTheoryBLS}.\par
The propagation of token distribution can be defined as a closed dynamical system where wealth moves between agents in exchange for goods and services. Equations in the dynamical system represent the propagation of wealth size of different agent categories with time.\par
We justify the use of the bilinear terms through previous works in econophysics, where it was empirically observed that wealth distribution in an economy relaxes to a specific wealth distribution \cite{RelaxationInStatisticalManyAgentEconomyModels}. In the case of standard economies, this distribution is the Pareto distribution with the parameter $\alpha\approx1.1$ \cite{HumanDevelopmentReport}. Not all economies necessarily satisfy this assumption, but empirical works on wealth distribution in token economies \cite{CharacterizingWealthInequality, HowCentralizedIsDecentralized} show that some economic metrics (i.e., Gini coefficient, Shannon entropy, Nakamoto index) applied to the wealth distribution demonstrate initial volatility of the wealth distribution that stagnates over time. Combining this stagnation in economic metrics in token economies with the relaxation  of wealth distribution in standard economies, we assume that token economies also experience relaxation in their wealth distribution with time.\par
The potential existence of equilibrium (i.e., the state the economy relaxes to) is what prompts us to use the bilinear form (i.e., matrix $\mathcal{B}$ in section \ref{DynamicalSystemGeneralizedSubsection}) in our equations, as it is the multi-dimensional version of the rate equation, which itself is concerned with the rate of interaction between specific compartments of the system and the attainment of an equilibrium by this system. Therefore, the motivation for using bilinear terms in our framework is that the real-world standard economies demonstrate equilibrium state (i.e., Pareto principle), and that token economies also empirically show the relaxation across different economic metrics (i.e., Gini coefficient, Shannon entropy, etc.).\par
\begin{remark}
	The discrete derivative of the wealth function for an agent category $A\in E_{t}$ with respect to time for every $t\in T$ (since $t_{initial}<t_{0}$) is defined by the difference quotient
	\begin{equation}
		\frac{\Delta F(A,t)}{\Delta t} = \frac{F(A,t)-F(A,t-\Delta t)}{\Delta t}.
	\end{equation}
\end{remark}
Consider a case of fungible tokens with two agent categories, $A,A^{\prime}\in E_{t}$, with wealth $F(A,t)$ and $F(A^{\prime},t)$ at time $t$ respectively. The wealth of $A$ evolves with
\begin{equation}
	\frac{\Delta F(A,t)}{\Delta t} = g(F(A,t),F(A^{\prime},t)),
\end{equation}
where $g$ is the function that represents redistribution of funds between $A$ and $A^{\prime}$ due to interactions and rotations. The propagation of wealth distribution of $A^{\prime}$ is
\begin{equation}
	\frac{\Delta F(A^{\prime},t)}{\Delta t} = g^{*}(F(A,t),F(A^{\prime},t)),
\end{equation}
where $g^{*}$ is also the function of token redistribution. If the maximum supply of tokens is capped by a constant value, $M$, it implies that
\begin{equation}
	M = F(A,t)+F(A^{\prime},t),
\end{equation}
and in its derivative form is
\begin{equation}
\label{DerivativeAssetInvariance}
	0 = \frac{\Delta F(A,t)}{\Delta t}+\frac{\Delta F(A^{\prime},t)}{\Delta t}.
\end{equation}
Cashflow interactions (transactions) are defined by
\begin{equation}
\label{C}
	C_{AA^{\prime}} = \beta_{AA^{\prime}}\frac{F(A,t)F(A^{\prime},t)}{M},
\end{equation}
where $C_{AA^{\prime}}$ is the wealth acquired by $A$ after transacting with $A^{\prime}$, and $\beta_{AA^{\prime}}$ is the interaction rate which is defined as the net wealth reallocated between $A$ and $A^{\prime}$ per wealth of $A$ per wealth of $A^{\prime}$ at timestep $t$ scaled by $M$.\par
\begin{remark}
	Translating the approach defined by Ross \cite{SIRModel} to our application will set $\beta$ as the number of contacts between agent $A$ and $A^{\prime}$ times the average probability of success of the interaction, however, here we set probability of success to 1 and get the deterministic distribution of wealth (the dimensionality of $\beta$ is $[time]^{-1}[wealth]^{-1}$).
\end{remark}
This type of interaction only represents the transactions between agent categories, it does not describe the rotation of participants from one agent category to another. The wealth rotations are defined as
\begin{equation}
\label{RA}
	R_{A} = \gamma_{A}F(A,t),
\end{equation}
where $R_{A}$ is the gross wealth that is redistributed due to the rotations of agents from the agent category $A$ to other agent categories, and $\gamma_{A}$ is the rotation rate of $A$.\par
For the equation \ref{RA} to be appropriately defined, the range is restricted to $\gamma_{A}\in[0,1]$ to prevent cases of negative wealth distribution and rotating more wealth than available.\par
The wealth rotation can proceed the opposite way with agents rotating from $A^{\prime}$ to $A$ via
\begin{equation}
\label{RAPrime}
	R_{A^{\prime}} = \gamma_{A^{\prime}}F(A^{\prime},t).
\end{equation}
Combining all the terms for transactions and rotations (equations \ref{C}, \ref{RA} and \ref{RAPrime}), the form of function $g$ is obtained,
\begin{equation}
	g(F(A,t),F(A^{\prime},t)) = \beta_{AA^{\prime}}\frac{F(A,t)F(A^{\prime},t)}{M}-\gamma_{A}F(A,t)+\gamma_{A^{\prime}}F(A^{\prime},t).
\end{equation}
From the equation \ref{DerivativeAssetInvariance}, and the fact that $M$ is constant
\begin{equation}
	g^{*}(F(A,t),F(A^{\prime},t)) = -\beta_{AA^{\prime}}\frac{F(A,t)F(A^{\prime},t)}{M}-\gamma_{A^{\prime}}F(A^{\prime},t)+\gamma_{A}F(A,t).
\end{equation}
Hence, the system of dynamical equations that models wealth distribution is
\begin{equation}
\label{DynamicalSystemMethodExample}
\begin{split}
	& \frac{\Delta F(A,t)}{\Delta t} = \beta_{AA^{\prime}}\frac{F(A,t)F(A^{\prime},t)}{M}-\gamma_{A}F(A,t)+\gamma_{A^{\prime}}F(A^{\prime},t) \\
	& \frac{\Delta F(A^{\prime},t)}{\Delta t} = -\beta_{AA^{\prime}}\frac{F(A,t)F(A^{\prime},t)}{M}-\gamma_{A^{\prime}}F(A^{\prime},t)+\gamma_{A}F(A,t).
\end{split}
\end{equation}
\subsection{Dynamical System (Generalized)}
\label{DynamicalSystemGeneralizedSubsection}
A dynamical system with $n$ agent categories can be constructed with the the same technique with no additional constraints introduced. The generalized form of the dynamical system for $n$ agent categories is:
\begin{equation}
\label{DynamicalSystemGeneralized}
	\frac{\Delta}{\Delta t}[\vv{F}(t)] = \frac{1}{M}\vv{F}(t)\odot[\mathcal{B}\cdot\vv{F}(t)]+\Gamma\cdot\vv{F}(t) \quad t\in T,
\end{equation}
where $\cdot$ is the matrix-vector product, $\odot$ is point-wise vector multiplication, $\vv{F}(t)$ is the vector of wealth functions at time $t$,
\begin{equation}
	\vv{F}(t) = (F(A_{1},t), ..., F(A_{n},t))^{T}, \quad A_{1},...,A_{n}\in E_{t},
\end{equation}
$\mathcal{B}$ is an antisymmetric matrix of interaction rates $\beta$,
\begin{equation}
\label{BetaMatrix}
	\mathcal{B} =
	\begin{bmatrix}
		0 & \beta_{A_{1}A_{2}} & \hdots & \beta_{A_{1}A_{n}} \\
		\beta_{A_{2}A_{1}} & 0 & \hdots & \beta_{A_{2}A_{n}} \\
		\vdots & \vdots & \ddots & \vdots \\
		\beta_{A_{n}A_{1}} & \beta_{A_{n}A_{2}} & \hdots & 0\\
	\end{bmatrix} =
	\begin{bmatrix}
		0 & \beta_{A_{1}A_{2}} & \hdots & \beta_{A_{1}A_{n}} \\
		-\beta_{A_{1}A_{2}} & 0 & \hdots & \beta_{A_{2}A_{n}} \\
		\vdots & \vdots & \ddots & \vdots \\
		-\beta_{A_{1}A_{n}} & -\beta_{A_{2}A_{n}} & \hdots & 0\\
	\end{bmatrix},
\end{equation}
and $\Gamma$ is the matrix of rotations where each column sums up to zero,
\begin{equation}
\label{GammaMatrix}
	\Gamma = 
	\begin{bmatrix}
		-\gamma_{A_{1}} & \gamma_{A_{2}A_{1}} & \hdots & \gamma_{A_{n}A_{1}} \\
		\gamma_{A_{1}A_{2}} & -\gamma_{A_{2}} & \hdots & \gamma_{A_{n}A_{2}} \\
		\vdots & \vdots & \ddots & \vdots \\
		\gamma_{A_{1}A_{n}} & \gamma_{A_{2}A_{n}} & \hdots & -\gamma_{A_{n}}\\
	\end{bmatrix},
\end{equation}
where the diagonal elements of $\Gamma$ are
\begin{equation}
	\gamma_{A_{m}} = \sum^{n}_{j\neq m}\gamma_{A_{m}A_{j}}.
\end{equation}
Note the difference between the rotation rates in the equation \ref{DynamicalSystemMethodExample} and the definition\ref{GammaMatrix}. Non-diagonal rotation rates from the definition \ref{GammaMatrix} have two indices. The reason behind this is that we need to identify the direction of the rotation rates and their sizes for economies that have more than two agent categories who can perform rotations (in general, $\gamma_{A_{j}A_{k}}\neq\gamma_{A_{k}A_{j}}$).
\begin{remark}
It is possible to introduce a game theory approach that models rotation terms in $\Gamma$ to imitate the generalized behaviour of agent categories but we will not expand on this notion in this paper.
\end{remark}
Depending on its structure, bilinear dynamical systems may exhibit stability and attain an attractor state \cite{OnTheConceptOfAttractor}, but in some cases dynamical systems may demonstrate mathematical chaos \cite{LorenzChaosTheory}. For a simple dynamical system, the proof of existence of an attractor and its analytic solution is relatively straight forward, however, for more complicated systems numerical solutions have to be employed.\par
An attractor represents no fluctuation in net wealth values across all agent categories and their wealth functions, but the interactions and rotations still take place and so does the wealth reallocation. We refer to this as the stability of the token economy. An attractor is defined by the following:
\begin{definition}
	\textbf{Attractor}, $x\in\mathbb{R}^{n}_{\geq0}$, of a dynamical system $h:\mathbb{R}^{n}_{\geq0}\times\mathbb{N}_{0}\rightarrow\mathbb{R}^{n}_{\geq0}$ is a point with a condition that $\exists t_{L}\in\mathbb{N}_{0}$ such that $\forall t\geq t_{L}$: $h(x,t)=x$ \cite{OnTheConceptOfAttractor}.
\end{definition}
If attractors in a token economy exist, the interaction and rotation rates are used to produce different attractors. For a given attractor, we can attempt to numerically solve the dynamical system to obtain the required constant parameters, but the solution may not always exist.
\subsection{Price Manifold and Hyperplane}
Through careful selection of interaction rates it is possible for the dynamical system to be directed to the desired token distribution or a numerically close to the desired distribution.\par
The constant interaction rate $\beta_{AA^{\prime}}$ in the equation \ref{C} can be written in terms of interaction quantities $\iota$ for different interaction types $i_{AA^{\prime}}\in I_{AA^{\prime}}$ at time $t$
\begin{equation}
\label{BetaAAPrime}
	\beta_{AA^{\prime}} = \frac{M}{\Delta t}\frac{\sum_{i_{AA^{\prime}}\in I_{AA^{\prime}}}\iota(i_{AA^{\prime}},t)}{F(A,t)F(A^{\prime},t)} \quad \forall t\in T,
\end{equation}
where individual $\iota$ is the wealth reallocated due to the specific interaction type $i_{AA^{\prime}}$ between agent categories $A$ and $A^{\prime}$ throughout the recent time interval $\Delta t$. Note that we can verify that the interaction rates are appropriately defined using the dimensional analysis \cite{DimensionalAnalysis}.\par
When using inverse propagation to find interaction and rotation rates that lead the wealth distribution dynamics towards the desired final state, we keep $\beta_{AA^{\prime}}$ constant in time. Constant parameters imply that interaction quantities $\iota$ for different interaction types $i_{AA^{\prime}}\in I_{AA^{\prime}}$ have to be rebalanced dynamically if the wealth of $A$ and $A^{\prime}$ changes throughout the recent $\Delta t$.\par
For a discrete time slice $t$ and labelling $H(t)=\frac{1}{M}\beta_{AA^{\prime}}(F(A,t)F(A^{\prime},t))$, we can rewrite the equation \ref{BetaAAPrime} as
\begin{equation}
\label{HyperplaneEquation}
	H(t) = \sum_{i_{AA^{\prime}}\in I_{AA^{\prime}}}\iota(i_{AA^{\prime}},t) = \langle\vv{D}(t),\vv{P}(t)\rangle,
\end{equation}
which is an equation of a hyperplane at every $t\in T$ for the constant vector of demands $\vv{D}(t)$, and the vector of prices $\vv{P}(t)$, where $\langle ,\rangle$ is the inner product and
\begin{equation}
\label{InteractionQuantity}
	D_{i_{AA^{\prime}}}(t)P_{i_{AA^{\prime}}}(t) \coloneqq \iota(i_{AA^{\prime}}, t).
\end{equation}
We justify this notation by stating that the sum of all interaction quantities is the sum of all transactions that took place in the recent $\Delta t$, and each transaction had a price and a quantity of goods or services that were exchanged at this price (prices can be negative implying the direction of the wealth transfer). If there were $n$ interaction types that took place, vectors $\vv{D}(t)$ and $\vv{P}(t)$ are both $n$-dimensional.\par
Given the transaction requests ($n$ demands) have been collected for the recent $\Delta t$, we can define a set of candidate price vectors that could be set in the token economy at time $t$. To keep the set of candidate price vectors and its constraints more general we define it through the notion of a manifold.\par
\begin{definition}
	\textbf{Set of price vectors}, $\mathcal{M}$, is a manifold, as defined by Carroll \cite{SpacetimeAndGeometry}, with each dimension constrained by the size of the current circulation.
\end{definition}
Note that the boundary of the price manifold cannot be larger than the current circulation $S(t)$. This implies that the set of price vectors is $\mathcal{M}\subset(-S(t),S(t))^{n}$ for $n$ demands and time $t\in T$. The boundary of the price manifold where $\vv{P}(t)=S(t)$ is not inside the manifold since $\mathcal{M}$ would not resemble the Euclidean space at that point.\par
Equation \ref{HyperplaneEquation} is the equation in the $n$-dimensional space of prices given demand vector $\vv{D}(t)$ is constant (it will be at $t$ as transaction requests are collected throughout $\Delta t$ and processed at discrete times $t\in T$). For a constant $\vv{D}(t)$, equation \ref{HyperplaneEquation} is the equation of a price hyperplane, and its intersection with the price manifold $P$ is the set of available price vector solutions that satisfy the conditions of being smaller than the current circulation and that they reconstruct the required $\beta_{AA^{\prime}}$.\par
This resembles the definition of demand curves in classical economics \cite{PrinciplesOfEconomics}. The volatility of demands at every $t$ varies, implying prices fluctuate as well. It is possible to restrict the prices for the interaction types (goods and services) that are more common in the token economy, but at an expense of a significant volatility in prices for the uncommon interaction types. Hence, it is possible to discriminate between assets based on their popularity among agents of token economies, and set up individual elasticities of demand for different types of interactions.
\subsection{Interlude: Relationship to Existing Distribution Models}
We previously made a remark that the interaction rate $\beta_{AA^{\prime}}$ can be written in a probabilistic way. This is useful in simulating the future of the wealth distribution based on the historical transaction data. In order to achieve this, we need to add the terms that represent the probability of success for a given interaction type to take place into the equation \ref{BetaAAPrime},
\begin{equation}
\label{BetaAAPrimeBernoulli}
	\beta_{AA^{\prime}} = \frac{M}{\Delta t}\frac{\sum_{i_{AA^{\prime}}\in I_{AA^{\prime}}}E[\mathcal{I}_{i_{AA^{\prime}}}]\iota(i_{AA^{\prime}}, t)}{F(A,t)F(A^{\prime}, t)} = \frac{M}{\Delta t}\frac{\sum_{i_{AA^{\prime}}\in I_{AA^{\prime}}}p_{i_{AA^{\prime}}}\iota(i_{AA^{\prime}}, t)}{F(A,t)F(A^{\prime}, t)}, \quad t\in T,
\end{equation}
where $\mathcal{I}_{i_{AA^{\prime}}}\sim Bernoulli(p_{i_{AA^{\prime}}})$ is the random variable describing the interaction type $i_{AA^{\prime}}$ taking place with a probability $p_{i_{AA^{\prime}}}$. Note, that by plugging the equation \ref{InteractionQuantity} into the equation \ref{BetaAAPrimeBernoulli}, we get
\begin{equation}
\label{BetaAAPrimeBinomial}
	\beta_{AA^{\prime}} = \frac{M}{\Delta t}\frac{\sum_{i_{AA^{\prime}}\in I_{AA^{\prime}}}p_{i_{AA^{\prime}}}D_{i_{AA^{\prime}}}(t)P_{i_{AA^{\prime}}}(t)}{F(A,t)F(A^{\prime}, t)} = \frac{M}{\Delta t}\frac{\sum_{i_{AA^{\prime}}\in I_{AA^{\prime}}}E[\tilde{\mathcal{I}}_{i_{AA^{\prime}}}]P_{i_{AA^{\prime}}}(t)}{F(A,t)F(A^{\prime}, t)},
\end{equation}
where $\tilde{\mathcal{I}}_{i_{AA^{\prime}}}\sim B(D_{i_{AA^{\prime}}}(t), p_{i_{AA^{\prime}}})$ is the binomial distribution for multiple interactions of type $i_{AA^{\prime}}$ to take place at a specific price level $P_{i_{AA^{\prime}}}(t)$ at time $t$. This is the probabilistic form of the equation \ref{BetaAAPrime} with the transaction requests being the random variables $\tilde{\mathcal{I}}_{i_{AA^{\prime}}}$ which can model different effects (e.g., success of transaction going through in unknown, forecasting wealth distribution based on mean transaction requests). The form of $\beta_{AA^{\prime}}$ in the equation \ref{BetaAAPrimeBinomial} is consistent with that defined by Ross \cite{SIRModel}, but we define it in the economics context.\par
In this paper we introduce the concept of economic taxonomy consisting of agent categories and interaction types, and propose modelling wealth distribution with constant parameters to find the prices in the economy. However, if we drop the concept of economic taxonomy and consider individual agents such that $\{a\}=A$ for $a\in \Lambda$ and $\vert E\vert=\vert\Lambda\vert$, drop $\gamma_{A}$ and $\gamma_{A^{\prime}}$ (since there are no more rotations as $\vert A\vert=1$), and set
\begin{equation}
	\beta_{AA^{\prime}}(t) = \frac{M}{\Delta t}\frac{\Delta F_{AA^{\prime}}(t)}{F(A,t)F(A^{\prime},t)},
\end{equation}
such that $\beta_{AA^{\prime}}$ becomes time dependent; with the help of the Euler method and setting $\Delta t=1$, we obtain the simplified version of the system of equations \ref{DynamicalSystemMethodExample},
\begin{equation}
\label{TradingRule}
\begin{split}
	& F(A,t+1) = F(A,t)+\Delta F_{AA^{\prime}}(t) \\
	& F(A^{\prime},t+1) = F(A^{\prime},t)-\Delta F_{AA^{\prime}}(t)
\end{split}
\end{equation}
which is the main trading rule set by Patriarca et al. \cite{KineticTheoryModelsForTheDistributionOfWealth} for wealth distribution in multi-agent models. Note that we can generalize this to $n$ agents using the general form of the dynamical system from the equation \ref{DynamicalSystemGeneralized} such that
\begin{equation}
	\frac{\Delta}{\Delta t}[\vv{F}(t)] = \frac{1}{M}\vv{F}(t)\odot[\mathcal{B}(t)\cdot\vv{F}(t)],
\end{equation}
with $\mathcal{B}$ defined as
\begin{equation}
	\mathcal{B}(t) =
	\begin{bmatrix}
		0 & \beta_{A_{1}A_{2}}(t) & \hdots & \beta_{A_{1}A_{n}}(t) \\
		\beta_{A_{2}A_{1}}(t) & 0 & \hdots & \beta_{A_{2}A_{n}}(t) \\
		\vdots & \vdots & \ddots & \vdots \\
		\beta_{A_{n}A_{1}}(t) & \beta_{A_{n}A_{2}}(t) & \hdots & 0\\
	\end{bmatrix}.
\end{equation}
Using the rule from the equation \ref{TradingRule} we can set $\Delta F_{AA^{\prime}}(t)$ to recreate the existing wealth distribution models as presented in Table \ref{WealthDistributionModels}. These distributions represent the relaxed states of the wealth distribution models for the given term $\Delta F_{AA^{\prime}}(t)$. The particularly interesting case is the wealth distribution model with individual saving propensities as it relaxes to the Pareto distribution with the parameter $\alpha\approx1.1$ \cite{RelaxationInStatisticalManyAgentEconomyModels}. Pareto distribution with this value of $\alpha$ fits the empirical observation, as per the United Nations Human Development Report \cite{HumanDevelopmentReport}, known as the Pareto principle that states that roughly 80\% of wealth belongs to the 20\% of population.\par
In these wealth distribution models, $\epsilon_{AA^{\prime}}$ is a random proportion of wealth redistributed between $A$ and $A^{\prime}$, with the complementary fraction $\bar{\epsilon}_{AA^{\prime}}=1-\epsilon_{AA^{\prime}}$. The saving propensity, $0<\lambda<1$, is the amount of wealth saved by each agent, while $\lambda_{A}$ is the saving propensity of agent $A$.\par
\begin{table}[h!]
\centering
\renewcommand{\arraystretch}{1.5}
\begin{tabular}{ |c|c| } 
\hline
$\Delta F_{AA^{\prime}}(t)$ & Wealth Distribution (Saving Mode) \\
\hline
$\bar{\epsilon}_{AA^{\prime}}F(A^{\prime},t)-\epsilon_{AA^{\prime}}F(A,t)$ & Boltzmann (No Saving) \cite{StatisticalMechanicsOfMoney} \\
$(1-\lambda)[\bar{\epsilon}_{AA^{\prime}}F(A^{\prime},t)-\epsilon_{AA^{\prime}}F(A,t)]$ & Gamma (Global Saving) \cite{StatisticalMechanicsOfMoneySavingPropensity} \\
$\bar{\epsilon}_{AA^{\prime}}(1-\lambda_{A^{\prime}})F(A^{\prime},t)-\epsilon_{AA^{\prime}}(1-\lambda_{A})F(A,t)$ & Pareto Tail (Individual Saving) \cite{MoneyInGasLikeMarkets} \\
\hline
\end{tabular}
\caption{Existing wealth distribution models.}
\label{WealthDistributionModels}
\end{table}
Note that our framework can be configured to implement the existing wealth distribution models. The relationship between our framework and the existing wealth distribution models demonstrates that our framework is flexible enough to be applied in different settings and that it is relevant to the research literature.

\section{Applications}
\label{Applications}
\subsection{Applications Scope}
To demonstrate possible applications of the framework from section \ref{MathematicalFramework} we will set up a simulation of a fictional economy. The reason we set up a simulation as opposed to try and find the solution in the closed form is because in general, bilinear dynamical systems may not have a solution, and for systems with more than two agent categories that do have solutions finding the exact analytical solutions may be very challenging and time-consuming.\par
Consider three agent categories - Consumer ($C$), Control Mechanism ($B$) and Producer ($P$), where Control Mechanism represents a mechanism that will be in control of monetary supply, fees and stimulus. Suppose that in our economy we will have four interaction types that lead to the reallocations of wealth between agent categories. The interaction types we use are defined in Table \ref{InteractionsTaxonomy}, along with the agent categories that will be spending or receiving wealth after each interaction.\par
\begin{table}[h!]
\centering
\begin{tabular}{ |c|c|c| } 
\hline
$i_{AA^{\prime}}$ (Interaction Type) & $A$ (Agent Category) & $A^{\prime}$ (Agent Category) \\
\hline
Good A & Consumer & Producer \\
Good B & Consumer & Producer \\
Maintenance Fee & Producer & Control Mechanism \\
Incentive & Control Mechanism & Consumer \\
\hline
\end{tabular}
\caption{Interactions taxonomy.}
\label{InteractionsTaxonomy}
\end{table}
Assume that in our economy there is a fixed maximum supply of tokens, $M$, and Control Mechanism is a standalone agent so nobody else in the economy can rotate into it, but this constraint aside, we will not be looking at individual agents or their microeconomic interactions in order to maintain macroeconomic scope of the simulation.\par
For the simulations, we will be using 400 timesteps so that $t_{0}=0$ and $t_{m}=399$ with $\Delta t=1$.\par
Next, let's assume that there exists a desired wealth distribution that optimizes a metric of choice, where the metric measures the benefit of allocating wealth in a specific way between agents in the economy. In the real-world setting this metric can be Gini Coefficient, GDP, etc. which will allow us to gauge what is the desired wealth distribution, but in our economy let's use the desired wealth distribution along with the initial wealths presented in Table \ref{InitialDesiredWealthDistribution}
\begin{table}[h!]
\centering
\begin{tabular}{ |c|c|c|c| } 
\hline
& Consumer & Control Mechanism & Producer \\
\hline
Initial Wealth & 1,500 & 98,000 & 500 \\
\hline
Desired Wealth & 50,000 & 10,000 & 40,000 \\
\hline
\end{tabular}
\caption{Initial and desired wealth distributions.}
\label{InitialDesiredWealthDistribution}
\end{table}
\begin{remark}
	Note that the number of points in the visualizations of the wealth distribution dynamics (Figures \ref{NoInteractions}, \ref{NoRotations}, \ref{ForwardPropagation} and \ref{InversePropagation}) are different across the charts. Throughout the study, the same number of points was used in every simulation, but some charts contain fewer points for display purposes alone.
\end{remark}
\subsection{Model}
We proceed with setting up a dynamical system using the technique from section \ref{MathematicalFramework}. For the three agent categories $C$, $B$ and $P$ we set up the equations of their wealth distribution dynamics,
\begin{equation}
\label{TestModel}
\begin{split}
	& \frac{\Delta F(P,t)}{\Delta t} = \beta_{PC}\frac{F(P,t)F(C,t)}{M}-\beta_{PB}\frac{F(P,t)F(B,t)}{M}-\gamma_{P}F(P,t)+\gamma_{C}F(C,t), \\
	& \frac{\Delta F(C,t)}{\Delta t} = \beta_{CB}\frac{F(C,t)F(B,t)}{M}-\beta_{PC}\frac{F(P,t)F(C,t)}{M}-\gamma_{C}F(C,t)+\gamma_{P}F(P,t), \\
	& \frac{\Delta F(B,t)}{\Delta t} = \beta_{PB}\frac{F(P,t)F(B,t)}{M}-\beta_{CB}\frac{F(C,t)F(B,t)}{M}.
\end{split}
\end{equation}
Given this dynamical system, we confirm that $M$ is conserved for every time $t$ by
\begin{equation}
	\frac{\Delta F(P,t)}{\Delta t}+\frac{\Delta F(C,t)}{\Delta t}+\frac{\Delta F(B,t)}{\Delta t} = 0.
\end{equation}
We further assume that the parameters $\beta_{PC}$, $\beta_{PB}$, $\beta_{CB}$, $\gamma_{P}$ and $\gamma_{C}$ are constant, and we leave the time-dependent interaction and rotation rates as well as the dynamic maximum supply $M(t)$ (e.g., maximum supply with incrementation) for future research.\par
In this example, the tokens are recycled as no tokens get burned. This may not be the case in the context of other token economies (e.g., Ethereum), where tokens are burned and can never return to the circulation as they become ``inaccessible'' to all agent categories, including the Control Mechanism. It is possible to add token burning by creating an agent category with one agent that acts as a token dump, where all burned tokens in the economy are allocated to. This agent must be unique and may not rotate into other agent categories (neither can other agents rotate into this ``token dump'' agent category), and it will have an additional restriction that it may not have a negative transaction quantity (i.e., it cannot send tokens or transact with the tokens that it holds). The inclusion of this agent category will add another term to the equation \ref{CirculationEquation} as the circulation will not include the wealth of this agent. To simplify the model we keep the token burning out of scope of this example.\par
Now we can apply finite difference method and simulate the wealth distribution dynamics with this system of equations using the forward propagation, or find the required interaction and rotation rates for the wealth dynamics to achieve the desired wealth distribution from Table \ref{InitialDesiredWealthDistribution} using the inverse propagation.
\subsection{Forward Propagation}
Forward propagation is a method for simulating wealth distribution in token economies based on the empirical data in the form of prices and demands (transaction requests). The objective of this procedure is to demonstrate the dynamics of interactions between agents and visualize trends in wealth redistribution.\par
\subsubsection{Inputs}
The inputs for the forward propagation are the interaction and rotation rates. In the context of our framework, the interaction rates (i.e., elements of matrix $\mathcal{B}$) represent the net wealth redistribution due to agents transacting with one another, while the rotation rates (i.e., elements of matrix $\Gamma$) are the gross wealth redistributed from one agent category to another due to individual agents changing their agent type affiliation to perform specific interactions limited to that agent category. These parameters are treated as constants throughout the simulation and will determine the dynamics of wealth distribution.\par
\begin{remark}
	When simulating wealth distribution in real-world economies we can use transaction data to compute the values of interaction and rotation rates. We can use the equation \ref{DynamicalSystemGeneralized}, where interaction and rotation rates are dynamic (i.e., $\mathcal{B}(t)$ and $\Gamma(t)$). The dynamic interaction rates can be obtained from the raw transaction data given the interactions taxonomy. For the dynamic rotation rates, we can inspect the transactions of agents to determine what agent categories the counter-parties of the transaction belong to and how they rotated between the agent categories historically. However, we will not expand on this notion in this paper.
\end{remark}
\subsubsection{Implementation}
We have defined the setting for our example economy and created a model for it, so now we can implement the forward propagation to demonstrate some characteristics of our framework, and attempt to find the parameters that force our model to propagate towards the desired wealth distribution.\par
Let us start with examining the impact that parameters of our model make on the dynamics of wealth distribution within our framework. As described in section \ref{DynamicalSystem}, interaction rates ($\beta_{PC}$, $\beta_{PB}$, $\beta_{CB}$) facilitate transactions in an economy, whereas rotation rates ($\gamma_{P}$, $\gamma_{C}$) represent the movement of wealth between agent categories due to individual agents changing their agent category affiliation (e.g., an agent earns wealth by manufacturing and selling goods as a Producer and decides to become a Consumer to spend wealth on goods himself).\par
\begin{figure}[h]
	\centering
	\includegraphics[scale=0.7]{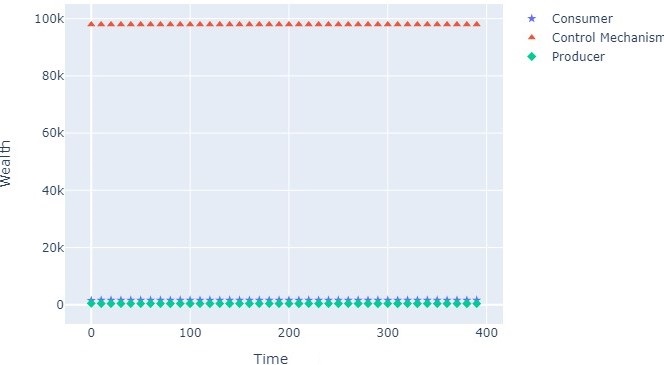}
	\caption{Wealth distribution dynamics without interactions.}
	\label{NoInteractions}
\end{figure}

\begin{table}[h!]
\centering
\begin{tabular}{ |c|c|c|c|c|c| } 
\hline
& $\beta_{PC}$ & $\beta_{PB}$ & $\beta_{CB}$ & $\gamma_{P}$ & $\gamma_{C}$\\
\hline
No Interactions & 0 & 0 & 0 & 0.2 & 0.05 \\
\hline
No Rotations & 0.3 & 0.2 & 0.35 & 0 & 0 \\
\hline
\end{tabular}
\caption{Parameters for extreme case simulations.}
\label{ExtremeCasesParameters}
\end{table}
We can simulate the economy in its extremes with either no interactions or no rotations. In the former case, we set $\beta_{PC}=\beta_{PB}=\beta_{CB}=0$, and set $\gamma_{P}=0.2$ and $\gamma_{C}=0.01$, as defined in Table \ref{ExtremeCasesParameters}. The reason for choosing these rotation rates is because we expect that the agent category with higher $\gamma$ value will eventually have more wealth, and we want to test this expectation. Plugging these parameters into the equation \ref{TestModel}, we see that the wealth of $B$ (Control Mechanism) will remain constant as $\frac{\Delta}{\Delta t}F(B,t)=0$, and the only wealth redistribution should be in the form of rotations between Consumer and Producer. Since $\gamma_{P}>\gamma_{C}$, we expect the Consumer to have more wealth by the end of the simulation, which is what we see on Figure \ref{NoInteractions} (``star'' markers are above ``rhombus'' markers).\par
\begin{figure}[h]
	\centering
	\includegraphics[scale=0.7]{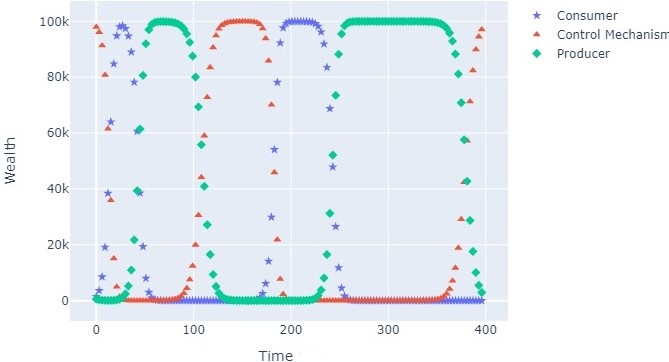}
	\caption{Wealth distribution dynamics without rotations.}
	\label{NoRotations}
\end{figure}
From Figure \ref{NoInteractions} we can see that indeed there is no wealth redistribution and hence no activity in the economy because all interaction rates are set to 0, which means there is no wealth redistribution as agents do not spend any wealth on purchasing goods and services.\par
The other extreme case scenario is the economy with no rotations. In this setting, we expect to see dynamical redistribution of wealth, but since agents are not allowed to rotate, there will be wealth hoarding by an agent category that is receiving wealth through interactions only (the wealth outflow from the hoarding agent category is limited by a constant interaction rate, creating a bottleneck effect in a cyclical system). For the simulation we set rotation rates to 0, and set $\beta_{PC}=0.3$, $\beta_{PB}=0.2$ and $\beta_{CB}=0.35$, where the choice of the interaction rates is arbitrary and is demonstrated in Table \ref{ExtremeCasesParameters}. The simulation of this case is presented on Figure \ref{NoRotations}.\par
Note the cyclical nature of the wealth distribution dynamics. This happens due to the way we have set up the interactions taxonomy in Table \ref{InteractionsTaxonomy}, as the agent type Consumer can only spend wealth by interacting with Producer, while Producer only spends wealth by interacting with Control Mechanism. This is a loop of wealth consumption in our closed economy which is prone to hoarding and lack of economic activity. Due to relatively high interaction rates we have set, the wealth redistribution happens in quick bursts (big proportion of total wealth moves in small amount of time), but we used these parameters to show how the interactions taxonomy impacts wealth distribution simulation in our framework.\par
Having explored examples of extreme scenarios and how the parametrization of our framework impacts the simulation outcomes, we now pick some arbitrary values for interaction and rotation rates and perform the forward propagation.\par
\begin{figure}[h]
	\centering
	\includegraphics[scale=0.7]{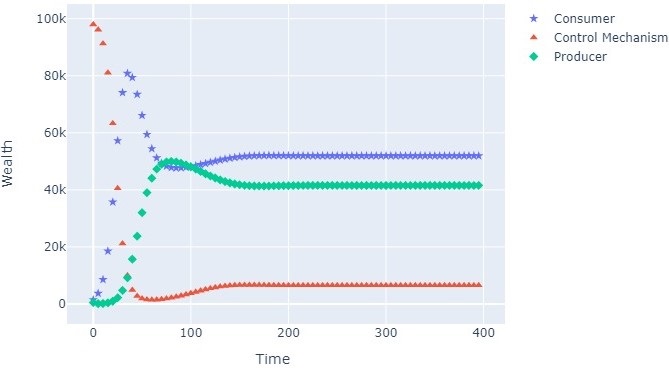}
	\caption{Forward propagation using parameters from Table \ref{ForwardPropagationParameters}.}
	\label{ForwardPropagation}
\end{figure}

\begin{table}[h!]
\centering
\begin{tabular}{ |c|c|c|c|c| } 
\hline
$\beta_{PC}$ & $\beta_{PB}$ & $\beta_{CB}$ & $\gamma_{P}$ & $\gamma_{C}$\\
\hline
0.2 & 0.25 & 0.2 & 0.1 & 0.01 \\
\hline
\end{tabular}
\caption{Arbitrary values of forward propagation parameters.}
\label{ForwardPropagationParameters}
\end{table}
Figure \ref{ForwardPropagation} shows the dynamics of wealth distribution in our economy using the parameters from Table \ref{ForwardPropagationParameters}. The dynamical system with our choice of parameters has converged to the final wealth distribution presented in Table \ref{ForwardPropagationFinalWealth}.\par
\begin{table}[h!]
\centering
\begin{tabular}{ |c|c|c| } 
\hline
Consumer & Control Mechanism & Producer \\
\hline
51,923 & 6,538 & 41,538 \\
\hline
\end{tabular}
\caption{Wealth distribution at time $t_{m}$ using parameters from Table \ref{ForwardPropagationParameters}.}
\label{ForwardPropagationFinalWealth}
\end{table}
The specific dynamics presented in Figure \ref{ForwardPropagation} has some features we can examine. First, the wealth is distributed from Control Mechanism to Consumer in the form of Incentive very quickly, which is due to the parameter $\beta_{CB}$ being set relatively high. Lowering this parameter will cause slower economic activity growth, as Consumer will spend less wealth on the goods offered by the Producer. This is an example of stimulus package being released into the economy to incentivize consumption.\par
Other feature we note on Figure \ref{ForwardPropagation} is the stabilization (constant wealth for each agent category) of wealth distribution at around $t=150$. This implies that the simulation has reached the attractor of the dynamical system and we will not see any further dynamics in the wealth distribution. This does not imply that there is no economic activity, instead it means that the current wealth distribution is balanced by the current economic activity. In the context of our economy, it means that Consumer spends as much wealth on products offered by the Producer as Consumer receives in incentives from the Control Mechanism.\par
To summarize, we have looked at an example of an application of our framework to modelling an economy with a predefined interactions taxonomy. We simulated the extreme cases that exist in our framework, understood why we are getting the specific dynamics, and how to interpret the parameters of the dynamical system.
\subsection{Inverse Propagation}
Inverse propagation is the method for finding the constant values of the interaction and rotation rates, motivated by identifying the correct parameter values that connect the initial and desired wealth distributions via some dynamic. Our methodology is to solve the equations \ref{TestModel} with the boundary conditions from Table \ref{InitialDesiredWealthDistribution} using the least squares algorithm.
\subsubsection{Inputs}
The inputs for the inverse propagation is the initial and the desired wealth distribution from Table \ref{InitialDesiredWealthDistribution} along with the interactions taxonomy from Table \ref{InteractionsTaxonomy}. Based on the interactions taxonomy our simulator initiates appropriate interaction rates (e.g., $\beta_{PC}$ relates to all possible interaction types between Consumer and Producer, so Good A and Good B).\par
\subsubsection{Implementation}
To implement the inverse propagation we use the assumption that if the set of interaction and rotation rates that achieve the desired wealth distribution exists, all derivatives in the equations \ref{TestModel} at the final timestep $t_{m}$ will be 0 as we aim to find an attractor of the dynamical system.\par
Setting the derivatives to 0, we can use the finite difference method to propagate the the dynamical equations with some constant interaction and rotation rates, just like in forward propagation, and subtract the desired wealths from the simulated final state of the economy. This gives us the difference between the simulated wealth distribution at $t_{m}$ and the desired wealth distribution. We can minimize this difference using the least squares solver from SciPy \cite{Scipy} to obtain the optimal set of parameters that enable the simulation to achieve the desired wealth distribution. Performing this procedure, we obtain the parameters in Table \ref{InversePropagationParameters}.\par
\begin{figure}[h]
	\centering
	\includegraphics[scale=0.7]{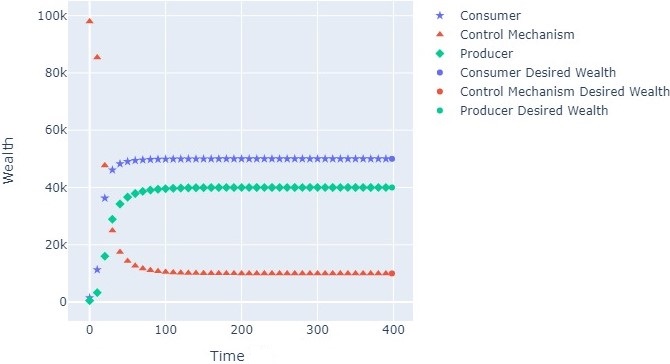}
	\caption{Inverse propagation using parameters from Table \ref{InversePropagationParameters}.}
	\label{InversePropagation}
\end{figure}

\begin{table}[h!]
\centering
\begin{tabular}{ |c|c|c|c|c| } 
\hline
$\beta_{PC}$ & $\beta_{PB}$ & $\beta_{CB}$ & $\gamma_{P}$ & $\gamma_{C}$\\
\hline
0.44265 & 0.56809 & 0.45447 & 0.63014 & 0.37250 \\
\hline
\end{tabular}
\caption{Solution of the inverse propagation.}
\label{InversePropagationParameters}
\end{table}
Having obtained the parameters we can plug them into the forward propagation simulation to find the dynamics of wealth distribution under these conditions. The simulation with new parameters is presented on Figure \ref{InversePropagation}.\par
Using the parameters from Table \ref{InversePropagationParameters} we achieved the desired wealth distribution from Table \ref{InitialDesiredWealthDistribution}. This method allows us to find the optimal parameters we need to set and from the equations \ref{BetaAAPrime} and \ref{HyperplaneEquation} we can directly set prices for different interaction types that exist within our interactions taxonomy.\par
By relating interaction types directly to the simulation, it is possible to introduce other methods for regulating and controlling the economy such as adding or removing interaction types, imposing quotas (e.g., maximum quantity of goods sold allowed), or setting inflation rate, but we leave these for future papers.\par
\begin{remark}
Note, that to run the inverse propagation the least squares solver from the SciPy library \cite{Scipy} was used. For more complex economic systems with a large number of agent categories this numerical solution may not be feasible. However, since the desired wealth distribution is known at the initial timestep, the dynamical system can be written in the form of Bellman equation and solved backwards \cite{DynamicProgramming}. For this, gradient descent or its variants such as stochastic gradient descent can be used, but we will leave the evaluation of the performance of these techniques with respect to our framework for future research.
\end{remark}
In summary, we have defined the inverse propagation method that allows us to compute the parameters required for our economy model to achieve the desired wealth distribution. The caveat of this approach is the fact that we assume that there exists an attractor for the dynamical system which can be reached in a finite time and that there exists a set of parameters that leads the wealth dynamics to this attractor.\par
\subsubsection{Price Solution}
We have mentioned the use of the equations \ref{BetaAAPrime} and \ref{HyperplaneEquation} for pricing interactions. In this section we demonstrate how using the constant interaction rates helps us with pricing of goods and services in our example economy.\par
From the interactions taxonomy defined in Table \ref{InteractionsTaxonomy}, we will price Good A and Good B using the interaction rate $\beta_{PC}$ that we found earlier. The reason we select $\beta_{PC}$ is because both, $\beta_{PB}$ and $\beta_{CB}$, have only one interaction associated to each of them (Maintenance Fee and Incentive respectively) as per Table \ref{InteractionsTaxonomy}, which makes these cases trivial with a unique price that solves the price hyperplane equation in each case.\par
Recall that a price hyperplane in the form of the equation \ref{HyperplaneEquation} is constructed at every timestep $t$, so let us choose an arbitrary timestep to demonstrate an example of the price hyperplane construction.\par
In our example we use the wealth dynamics presented in Figure \ref{InversePropagation} at time $t=50$ with wealths of agent categories at $t$ listed in Table \ref{PricingExampleWealths}. We also need to define a vector of demands, $\vv{D}(t)$, which represents transaction requests for different interaction types from the interactions taxonomy that take place between the agent categories. In this example, we have Good A and Good B that are related to $\beta_{PC}$ (as these are the only interaction types that can occur between Producer and Consumer in our taxonomy) so $\vv{D}(t)$ is the vector of transaction requests that have been collected in the latest $\Delta t$. Let us assume that in the latest $\Delta t$ (the period between times $t=49$ and $t=50$) Consumer has submitted 30 buy orders for Good A and 60 buy orders for good B, therefore, we have that $\vv{D}(50)=(30, 60)^{T}$.\par
\begin{figure}[h]
	\centering
	\includegraphics[scale=0.7]{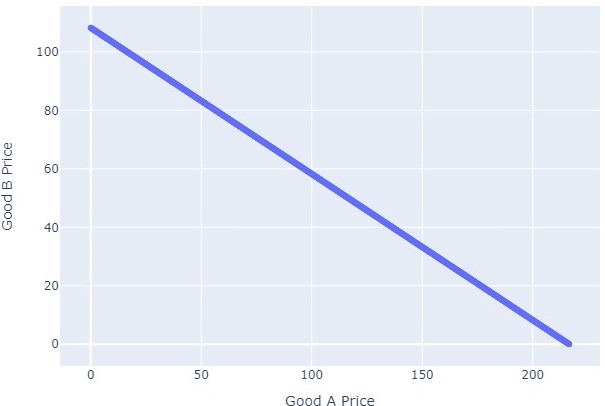}
	\caption{Price hyperplane constructed from the equation \ref{PricingExampleHyperplane}.}
	\label{PricingExampleFigure}
\end{figure}

\begin{table}[h!]
\centering
\begin{tabular}{ |c|c|c| } 
\hline
Consumer & Control Mechanism & Producer \\
\hline
49,053 & 14,309 & 36,638 \\
\hline
\end{tabular}
\caption{Wealths of agent categories at timestep $t=50$ using parameters from Table \ref{InversePropagationParameters}.}
\label{PricingExampleWealths}
\end{table}
Using the equation \ref{HyperplaneEquation} we construct the general price hyperplane equation in terms of $\beta_{PC}$,
\begin{equation}
\label{PricingExampleGeneralHyperplane}
	\frac{1}{M}\beta_{PC}F(P,t)F(C,t) = \langle\vv{D}(t),\vv{P}(t)\rangle
\end{equation}
and plug in in the values at $t=50$ so that we obtain the expression for the price hyperplane at this timestep,
\begin{equation}
\label{PricingExampleHyperplane}
	6496 = 30P_{\text{Good A}}(50)+60P_{\text{Good B}}(50),
\end{equation}
where $P_{\text{Good A}}(50)$ and $P_{\text{Good B}}(50)$ are the prices of Good A and Good B respectively at timestep $t=50$. The prices that solve this equation will satisfy the interaction rate $\beta_{PC}=0.44265$ (i.e., the $\beta_{PC}$ from Table \ref{InversePropagationParameters}) that we need to reach the desired wealth distribution. Figure \ref{PricingExampleFigure} presents the price hyperplane from the equation \ref{PricingExampleHyperplane}.\par
\begin{remark}
	Note that we constraint our demand vector to be a non-zero vector. The reason for this limitation is to ensure $\beta_{PC}$ is constant and non-zero. In macroeconomic setting this assumption is justified by the scale of the economy, where we can always assume at least some goods have been purchased in the recent time interval, however, this assumption will not be true if we consider a case where agent categories are individual agents instead of being sets of agents. In that case we will have to use dynamic interaction and rotation rates.
\end{remark}
Note the relationship between the prices for Good A and Good B on Figure \ref{PricingExampleFigure}. The inverse relationship between the prices is the result of our proposed balancing technique to obtain the desired $\beta_{PC}$. Any point located on the line will give us the desired $\beta_{PC}$ at $t=50$ with the demand vector $\vv{D}(50)=(30,60)^{T}$.\par
In this case, the price hyperplane is a 1-dimensional line embedded in a 2-dimensional Euclidean space. If we were to add another interaction type that can occur between Producer and Consumer to the  interactions taxonomy, we would obtain a 2-dimensional hyperplane embedded in a 3-dimensional space.\par
\begin{remark}
	Note that we did not present the equations that define the rotation rates in terms of empirical data. The motivation behind this is that when employing inverse propagation, in most cases the rotation rates will not be constant as rotations are motivated entirely by agents' behaviour and wealth consumption preferences. However, in this scenario we can solve the equations \ref{TestModel} for the interaction and rotation rates at every timestep (using wealth distribution from the previous timestep as initial conditions) and employ the same pricing mechanism for the interaction rates as outlined in this section.
\end{remark}
To summarize, we presented an approach to interpret the interaction rates from our simulation in terms of empirical data (transactions), and found a way to maintain the needed interaction rates to achieve the desired wealth distribution through the selection of prices. If prices in the economy are set by agents themselves, we can introduce taxes or stimulus as interaction types where their values are set by the Control Mechanism, and these interaction types act as a balancing mechanism in the economy (i.e., they help us obtain the desired interaction rates from their respective price hyperplanes). The purpose of the pricing method outlined here is to set the prices that achieve a desired wealth distribution, or to act as an advisory tool for the token economy governors and regulators to make policy decisions with the extra quantitative support.\par
\subsection{Interlude: Pricing Solution Extension}
In the previous section we studied the theoretical price solution that can be obtained based on the inverse propagation. The price solution tells us what prices must be set in the token economy for it to achieve a desired wealth distribution, but it relies on the assumption that all prices are set by the Control Mechanism. If the token economy in question is the economy where all prices are set by the Control Mechanism, then the pricing solution will give the desired dynamic of wealth distribution. However, if the Producer is allowed to set his own prices, he may not set the prices that will produce the value of $\beta_{PC}$ that we want.\par
This problem is solved by adding interactions between the Producer and Consumer, such that the Control Mechanism is in charge of setting the sizes of these interactions. Given the previous example, consider adding interaction types called ``Stability Tax For Good A'' and ``Stability Tax For Good B'', where the Control Mechanism sets their ``prices'' directly. The objective of these new interactions is to rebalance the interaction quantities $\iota$ between Producer and Consumer at time $t$, which allows us to obtain the $\beta_{PC}$ we need for the wealth distribution to behave according to the inverse propagation solution we found.\par
Let the prices $P_{\text{Good A}}(t)$ and $P_{\text{Good B}}(t)$ be set by the Producer and require these prices to be known at timestep $t$ (i.e., the prices can be set by the Producer at any timestep before the current timestep $t$) and they have to be constant in the latest interval $\Delta t$.\par
Recall the equation \ref{PricingExampleGeneralHyperplane},
\begin{equation}
	\frac{1}{M}\beta_{PC}F(P,t)F(C,t) = \langle\vv{D}(t),\vv{P}(t)\rangle,
\end{equation}
that we used to construct the pricing hyperplane at time $t=50$, which is given in the equation \ref{PricingExampleHyperplane}. Without loss of generality, assume some arbitrary prices for Good A and Good B were set by the Producer so that at time $t=50$ these prices are quoted as $P_{\text{Good A}}(50) = 100$ and $P_{\text{Good B}}(50) = 80$. Note that these prices do not lie on the price hyperplane demonstrated on Figure \ref{PricingExampleFigure}, and therefore, do not satisfy the equation \ref{PricingExampleHyperplane}.\par
For each price we introduced a new interaction type in form of stability tax, which we can add into the price hyperplane equation with the same demand as the interaction we impose stability tax on (i.e., the demand for Stability Tax For Good A at time $t$ is the same as the demand for the Good A) such that the price hyperplane becomes
\begin{equation}
\begin{split}
	6496 =& 30(P_{\text{Good A}}(50)+P_{\text{Stability Tax For Good A}}(50))+\\
	&+60(P_{\text{Good B}}(50)+P_{\text{Stability Tax For Good B}}(50)),
\end{split}
\end{equation}
where we can plug in the prices set by the Producer to obtain
\begin{equation}
\begin{split}
	6496 = 30(100+P_{\text{Stability Tax For Good A}}(50))+60(80+P_{\text{Stability Tax For Good B}}(50)),
\end{split}
\end{equation}
so that instead of the prices, we can find the values of the stability taxes to rebalance the transactions between the Producer and Consumer.\par
This notion can be generalized beyond this example. It is achieved by adding new interactions, whose interaction quantity $\iota$ is determined by the Control Mechanism, for every ``free'' interaction in $I_{AA^{\prime}}$ between agent categories $A\in E_{t}$ and $A^{\prime}\in E_{t}$. This allows the Control Mechanism to rebalance the cashflow interactions between these two agent categories to achieve the desired interaction rate $\beta_{AA^{\prime}}$.\par
Note that in the example above, it may be required for  the Consumer to pay more for the good than the quoted price, if the quoted prices (i.e., the coordinate ($P_{\text{Good A}}(50)$, $P_{\text{Good B}}(50)$) set by the Producer) are below the price hyperplane demonstrated on Figure \ref{PricingExampleFigure}. The opposite can also be true, where the Producer might be asked to reimburse the Consumer if the prices quoted are above the price hyperplane.\par
This also means that regardless of the price set by the Producer, the Producer will still generate the same revenue, which is the limitation of this pricing approach. However, it can be argued that if the economy designers require the economy to act according to a specific wealth distribution dynamic, why allow agent categories to quote the prices in the first place? This is the question on the degree of the control that exists in an economy that extends beyond our framework and is debated by economists for decades, so will not be addressing it in this paper, rather we just state the method to price goods to achieve the desired wealth distribution where the Control Mechanism has the ability to do so.\par

\section{Conclusion}
Our objective was to design a new framework that models token economies, performs pricing and introduces regulatory control mechanism. With DeTEcT we constructed the framework to model the wealth distribution dynamics and we introduced a new way to analyse the economic activity through low-level transactions and interactions taxonomy. If a token economy has an attractor (e.g., desired wealth distribution), using our framework it is possible to find the parameters of the token economy that will demonstrate the convergence, and price interactions such as goods, fees, incentives, etc. In the future research, we aim to extend this framework to include features like monetary supply incrementation, economic metrics, and other tools for economic analysis. Moreover, we would like to explore ways of controlling convergence rate of the dynamical system and incorporate it into the inverse propagation simulation to yield parameters that will perform wealth redistribution at a slower rate with less volatile economic dynamics.\par
The framework we have presented in this paper is designed to be implemented in token economies as these have full transaction data available on the ledgers. Our main objective for the future research is to implement this framework to model a real-world token economy such as the economy of the Ethereum network, and understand how control mechanisms such as money supply, fees, or incentives, can be implemented to produce a desired wealth distribution (according to a metric of choice).\par
The idea behind DeTEcT is to have a framework for designing token economies with long-term economic stability, algorithmic policy implementation, and a price-setting mechanism, while leaving the flexibility for the economy to change its interactions taxonomy or desired objectives (i.e., metrics of ``success''). We believe that with such strong set of applications, DeTEcT is a successful formal analysis framework for wealth distribution simulation, and analysis of algorithmic policy implementation in token economies.

\section*{Acknowledgements}
We would like to thank the Reviewers and Editor of \emph{Frontiers in Blockchain} for their detailed and helpful feedback that contributed to improving this paper.

\end{document}